\documentclass[10pt,twocolumn,reqno, aps,prx,superscriptaddress]{revtex4-2}%
\usepackage{amsmath}
\usepackage[pdftex]{graphicx}
\usepackage{float}
\usepackage{color}
\usepackage{rotating}
\usepackage[breaklinks=true,colorlinks,citecolor=blue,linkcolor=blue,urlcolor=blue]%
{hyperref}
\usepackage{graphicx}
\usepackage{natbib}
\usepackage{dcolumn}
\usepackage{bm}
\usepackage{hyperref}
\usepackage[caption=false]{subfig}
\usepackage{epstopdf}
\usepackage{ulem}
\usepackage{amsfonts}
\usepackage{amssymb}%
\providecommand{\U}[1]{\protect\rule{.1in}{.1in}}
\providecommand{\U}[1]{\protect\rule{.1in}{.1in}}
\providecommand{\U}[1]{\protect\rule{.1in}{.1in}}
\begin{document}

\title{Nonlinear enhancement of coherent magnetization dynamics}
\author{Mehrdad Elyasi}
\affiliation{Advanced Institute for Materials Research, Tohoku University, Sendai 980-8577, Japan}
\author{Kei Yamamoto}
\affiliation{Advanced Science Research Center, Japan Atomic Energy Agency, Tokai 319-1195, Japan}
\author{Tomosato Hioki}
\affiliation{Department of Applied Physics, The University of Tokyo, Tokyo 113-8656, Japan}
\author{Takahiko Makiuchi}
\affiliation{Department of Applied Physics, The University of Tokyo, Tokyo 113-8656, Japan}
\affiliation{Quantum-Phase Electronics Center, The University of Tokyo, Tokyo 113-8656, Japan}
\author{Hiroki Shimizu}
\affiliation{Department of Applied Physics, The University of Tokyo, Tokyo 113-8656, Japan}
\author{ Eiji Saitoh}
\affiliation{WPI Advanced Institute for Materials Research, Tohoku University, Sendai 980-8577, Japan}
\affiliation{Department of Applied Physics, The University of Tokyo, Tokyo 113-8656, Japan}
\affiliation{Institute for AI and Beyond, The University of Tokyo, Tokyo 113-8656, Japan}
\affiliation{Advanced Science Research Center, Japan Atomic Energy Agency, Tokai 319-1195, Japan}
\author{Gerrit E. W. Bauer}
\affiliation{WPI Advanced Institute for Materials Research, Tohoku University, Sendai 980-8577, Japan}
\affiliation{Kavli Institute for Theoretical Sciences, University of the Chinese Academy of Sciences, Beijing 10090, China}
\affiliation{Zernike Institute for Advanced Materials, University of Groningen, 9747 AG Groningen, Netherlands}

\begin{abstract}
Magnets are interesting materials for classical and quantum information
technologies. However, the short decoherence and dephasing times that
determine the scale and speed of information networks, severely limit the
appeal of employing the ferromagnetic resonance. Here we show that the
lifetime and coherence of the uniform Kittel mode can be enhanced by 3-magnon
interaction-induced mixing with the long-lived magnons at the minima of the
dispersion relation. Analytical and numerical calculations based on this model
explain recent experimental results and predict experimental signatures of
quantum coherence.

\end{abstract}
\maketitle


Quantum information processing relies on long coherence times of the elements
in the network. These times can be enhanced by the controlled coupling of the
information carrier to other long-lived degrees of freedom in hybrid devices
\cite{Xiang2013}. For example, a collection of nitrogen-vacancy centers in
diamond can act as a memory for cavity photons with lower quality
\cite{Putz2014,Putz2016,Krimer2015,Krimer2014,Julsgaard2013}. Spin refocusing
and dynamic decoupling of qubits from their environment are other schemes to
tackle \textquotedblleft black box dissipation/dephasing
sources\textquotedblright\ \cite{Viola1998,Lange2010}.

The uniform magnetization oscillation or Kittel magnon strongly interacts
with microwave photons and be used to coherently control quantum states
\cite{Lachance2020,Elyasi2020,Sharma2021,Kounalakis2022,Yuan2022}, and be read
out electrically \cite{Hioki2021}. However, its lifetime is often shorter than
expected from the intrinsic Gilbert damping of yttrium iron garnet (YIG) with
reported quality factors as large as $10^{5}$ \cite{Stancil2009}. To
realize, implement, and assess recent proposals for the generation of
Kittel mode quantum states and their implementation in novel
computational paradigms \cite{Chumak2022,Rameshti2022,Yuan2022}, the Kittel
mode coherence must be pushed to its limits.

Magnon interactions have been mainly associated with an increase in
dissipation or dephasing, which can be useful for the control of the magnon
transport \cite{Kurebayashi2011,Baruskov2019,Divinskiy2019,Mohseni2021}.
However, three-magnon interactions may also increase lifetimes. According to
recent experiments, the Kittel mode lives longer by the coherent mixing with
spin waves that have finite wave vectors
\cite{Demidov2007,Demidov2008,Serga2013,Bozhko2016,Makiuchi2023}. We observed
an excitation power-dependent enhanced lifetime of the Kittel mode or
``persistent" coherence \cite{Makiuchi2023} and attributed it to three magnon
scattering (3MS). Similar physics also explains the power-dependent quenching
of the magnon-photon interaction \cite{Lee2023}. Here, we demonstrate how the
valley magnons at the minima of the dispersion relation enhance the coherence
lifetime of the Kittel mode in a monolithic magnet.

\begin{figure}[ptb]
\includegraphics[width=0.5\textwidth]{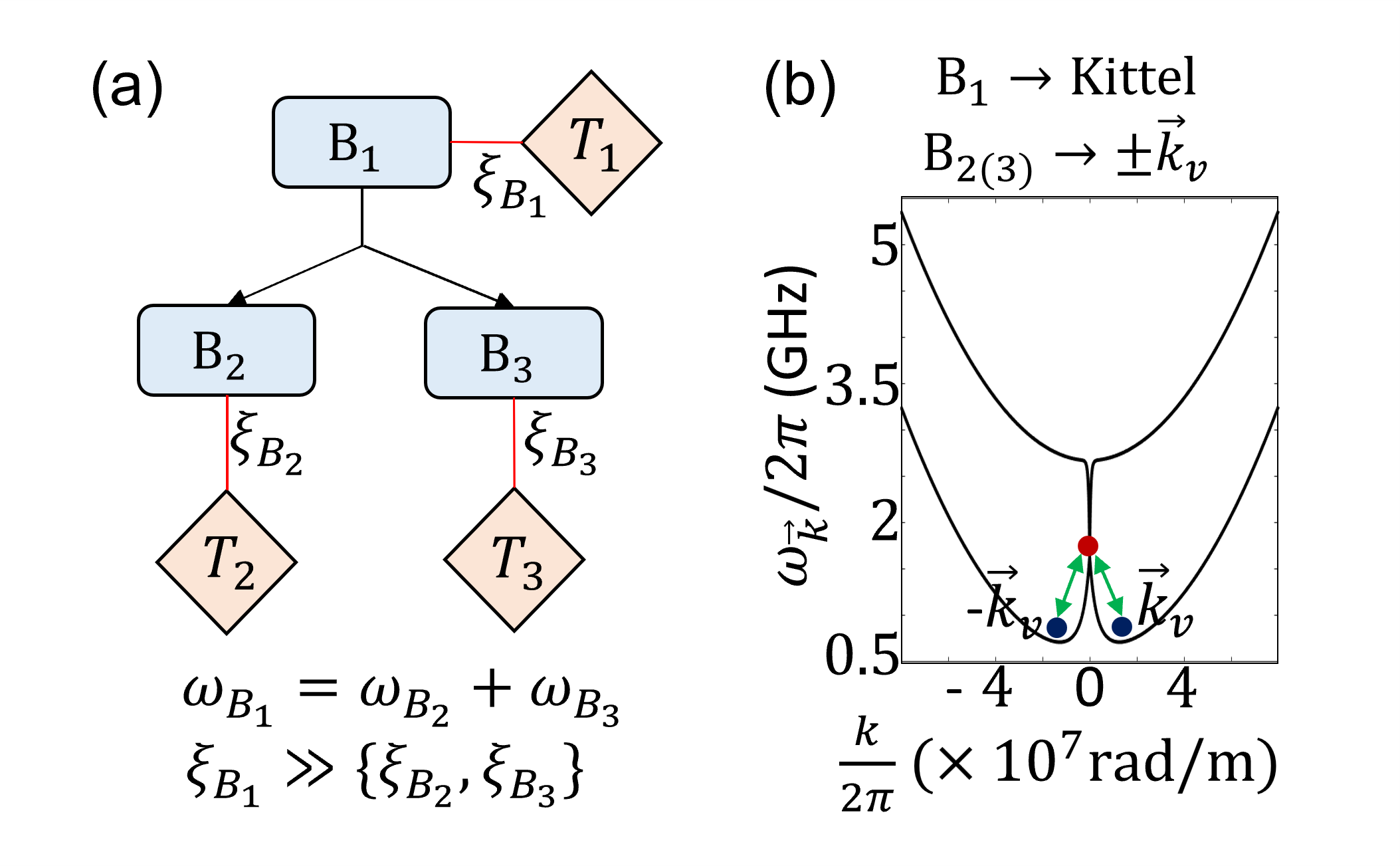}\caption{(a) Schematics of
the generic 3 state model with frequencies $\omega$ and dissipation rates
$\xi$. (b) In a thin film magnet, the Kittel mode interacts with valley magnon
pairs with wave vectors $\pm\vec{k}_{V}$. Here we show the dispersion
$\omega_{\vec{k}}$ of a $d=1\,\mathrm{\mu}$m YIG film, for $\vec{k}\Vert
\vec{H}$ (bottom curve) and $\vec{k}\perp\vec{H}$ (top curve), where $\vec{H}$
is an in-plane magnetic field.}%
\label{fig1}%
\end{figure}

\textit{Model-} Our generic model consists of three interacting Bosonic modes
$B_{1(2,3)}$ at (quasi-)equilibrium temperatures $T_{B_{1(2,3)}}$, and
dissipation rates $\xi_{B_{1(2,3)}}$ with frequencies that satisfy
$\omega_{B_{1}}=\omega_{B_{2}}+\omega_{B_{3}}$ and $\xi_{B_{1}}\gg\{\xi
_{B_{2}},\xi_{B_{3}}\}$ [see Fig. \ref{fig1} (a)]. We specialize in magnetic
thin films with a sufficiently strong magnetodipolar coupling that generates a
pronounced minimum frequency at finite wave vectors along the magnetization
direction (valley magnons) [see Fig. \ref{fig1}(b)], such that $B_{1}\ $is the
$\vec{k}=0$ Kittel mode with frequency $\omega_{0}$, $B_{2(3)}$ is a valley
magnon with momentum $+(-)\vec{k}_{V}$. The dispersion relations of magnetic
disks agree with those of extended films of the same thickness as long as the
demagnetizing fields are approximately constant across the sample and the
frequency spacing of standing waves does not exceed their inverse lifetime
\cite{Kalinikos1986,Kalinikos1990,Hurben1995}.

The magnetodipolar interaction becomes weaker with film thickness, with
shallower valleys and a minimum frequency that exceeds $\sim\omega_{0}/2,$
which suppresses the 3-magnon scattering for YIG films with thickness
$d\lesssim800\,$nm.
However, in small disks the inhomogeneous demagnetizing field at the edges
pulls the frequencies below the valley minimum $\omega_{min}$ with
corresponding amplitudes localized at the edges. The 3MS interaction of these
edge states with the Kittel magnons is therefore efficient even for thinner
films \cite{Makiuchi2023}.

The Lindblad master equation for the density matrix $\rho$ of the driven
interacting magnon gas reads
\cite{Carmichael1,Carmichael2,Krivosik2010,Elyasi2022},
\begin{align}
\dot{\rho}  &  =-i[\mathcal{H}^{(L)}+\mathcal{H}^{(NL)}+\mathcal{H}^{(d)}%
,\rho]+\nonumber\\
&  \sum_{\vec{k}}\xi_{\vec{k}}L_{\vec{k}}^{(L)}[\rho]+\sum_{\vec{k}}g_{\vec
{k}}L_{\vec{k}}^{(NL)}[\rho]. \label{eq1}%
\end{align}
where $\vec{k}\in\{0,\vec{k}_{V}\}$. The system is tuned such that the valley
magnon frequency is close to twice that of the Kittel mode $\omega_{\vec
{k}_{V}}\sim{\omega_{0}}/2$. In $\mathcal{H}^{(L)}=\sum_{\vec{k}}\Delta
\omega_{\vec{k}}c_{\vec{k}}^{\dag}c_{\vec{k}}$, $c_{\vec{k}}$ is the
annihilation operator of $\vec{k}$ magnon in a rotating frame so that $\Delta\omega_{0}=0$ and
$\Delta\omega_{k_{V}}=\omega_{k_{V}}-\omega_{0}/2$. The leading magnon
nonlinearities are $\mathcal{H}^{(NL)}=\mathcal{H}^{(3MS)}+\mathcal{H}%
^{(4MS)}$, where $\mathcal{H}^{(3MS)}=\mathcal{D}_{\vec{k}_{V}}c_{0}^{\dag
}c_{\vec{k}_{V}}c_{-\vec{k}_{V}}+\mathrm{H.c}.$ is the three-magnon
interaction with coefficient $\mathcal{D}_{\vec{k}_{V}}$, while $\mathcal{H}%
^{(4MS)}$ includes the four-magnon scattering terms
\cite{Krivosik2010,Elyasi2022}. The Kittel mode can be driven either
resonantly or parametrically, i.e. $\mathcal{H}^{(d)}=\left(  P_{r}c_{0}%
+P_{r}^{\ast}c_{0}^{\dag}\right)  +\left(  P_{p}c_{0}^{2}+P_{p}^{\ast}%
c_{0}^{\dag2}\right)  $, with amplitudes $P_{r}$ and $P_{p}$, respectively,
that are proportional to the power of an external drive. $\xi_{\vec{k}}$ is a
dissipation rate and $L_{\vec{k}}^{(L)}$ is a linear Lindblad dissipation
operator that acts only in the Hilbert space of the $\vec{k}$-mode. The
nonlinear dissipation operator $L_{\vec{k}}^{(NL)}$ and rate $g_{\vec{k}}$ are
the result of integrating out the Bosonic (photon or spin transfer) drive that
parametrically excites mode $\vec{k}$ \cite{Kinsler1991}. $L_{\vec{k}}%
^{(X)}=(\bar{n}_{\vec{k}}+1)(2c_{\vec{k}}^{q}\rho c_{\vec{k}}^{\dag q}%
-c_{\vec{k}}^{\dag q}c_{\vec{k}}^{q}\rho-\rho c_{\vec{k}}^{\dag q}c_{\vec{k}%
}^{q})+\bar{n}_{\vec{k}}(2c_{\vec{k}}^{\dag q}\rho c_{\vec{k}}^{q}-c_{\vec{k}%
}^{q}c_{\vec{k}}^{\dag q}\rho-\rho c_{\vec{k}}^{q}c_{\vec{k}}^{\dag q}),$
where $X\in\{L,NL\}$, $q=\delta_{X,L}+2\delta_{X,NL}$, and $\bar{n}_{\vec{k}}$
is the average particle number of a thermal Bosonic bath formed by other
magnons and/or phonons that at equilibrium reduces to a Planck distribution
at temperature $T$. In this work, we focus on the resonant drive of the Kittel mode and adopt $P_{p}=g_{\vec{k}}=0$ throughout, but all our findings on the decay rates remain valid when the drive is parametric \cite{Makiuchi2023}.

\begin{figure}[ptb]
\includegraphics[width=0.5\textwidth]{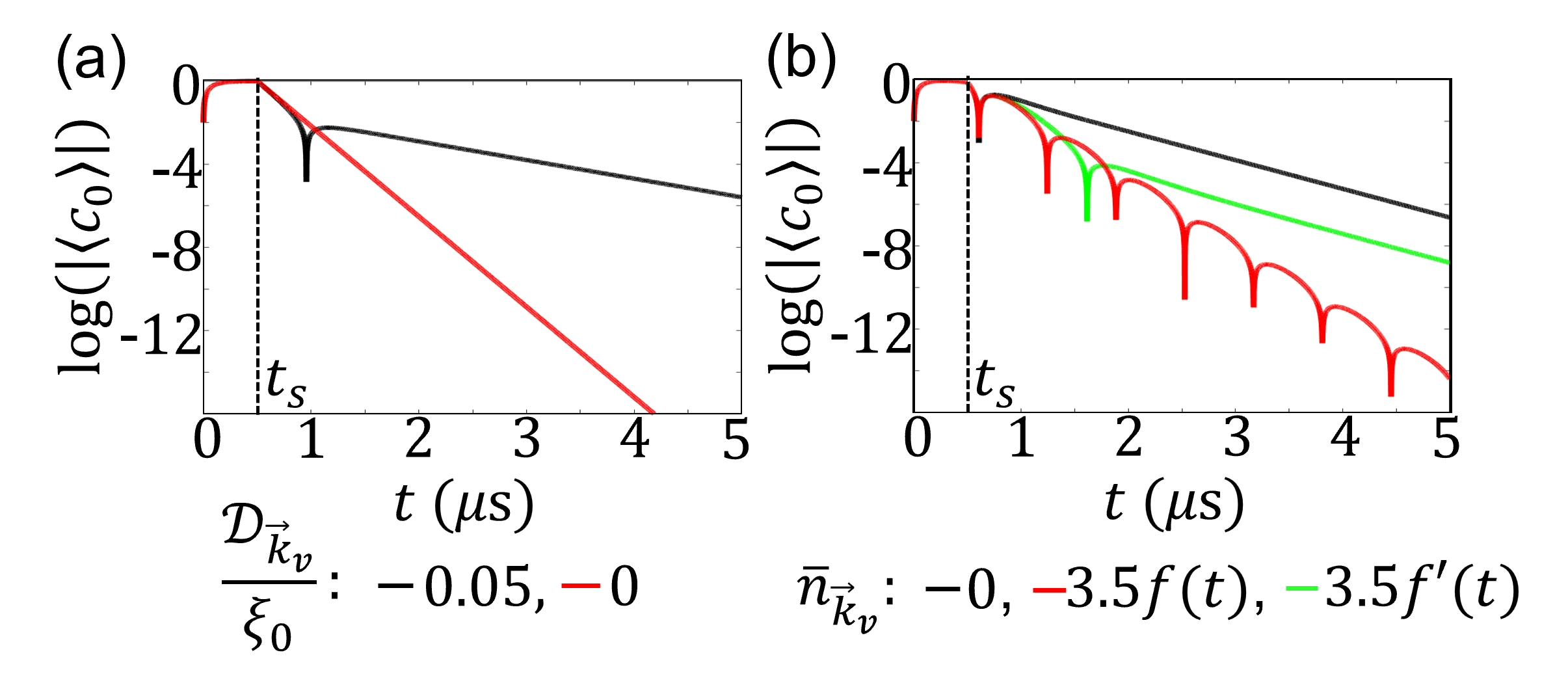}\caption{Numerical
calculation of the Kittel mode decay in the presence of a single valley magnon
pair mode under resonant excitation $P_{r}=0.5\xi_{0}[\Theta(t)-\Theta
(t-t_{s})]$, $\xi_{0}=1\,$MHz, $\xi_{\vec{k}_{V}}=0.1\,$MHz, $\mathcal{H}%
^{(4MS)}=0$. (a) $\langle c_{0}\rangle(t)$ for two
values of $\mathcal{D}_{\vec{k}_{V}}/\xi_{0}=0.05$ and $0$ and zero
temperature $n_{th,0(\pm\vec{k}_{V})}=0$. (b) $\langle c_{0}\rangle(t)$ for
three different $n_{th,\pm\vec{k}_{V}}=0$, $3.5f(t)$ and $3.5f^{\prime}(t)$,
where $f(t)=\Theta(t-t_{s})$ and $f^{\prime}(t)=\Theta(t-t_{s})\exp
{[-2\xi_{\pm\vec{k}_{V}}(t-t_{s})]}$, for a stronger interaction
$\mathcal{D}_{\vec{k}_{V}}/\xi_{0}=0.25$.}%
\label{fig2}%
\end{figure}

In the following, we divide our attention among `classical' and `quantum'
coherence. Under classical coherence, we sort the lifetime of phase and
amplitude of coherent states, whereas quantum coherence describes the interference
of macroscopic wave functions formed by the superposition of coherent
states or \textquotedblleft Schr\"{o}dinger cat states\textquotedblright.

\textit{Classical coherence-} The effect of three-magnon scattering on
classical coherence of the Kittel mode is at the hand of the mean-field
amplitudes $\langle c_{0}\rangle$ and $\langle c_{-\vec{k}_{V}}c_{\vec{k}_{V}%
}\rangle$ with a coupled equation of motion (EOM) that follows from the master
Eq. (\ref{eq1}),
\begin{align}
&  \frac{d\left\langle c_{0}\right\rangle }{dt}=-i\sum_{\vec{k}_{V}%
}\mathcal{D}_{\vec{k}_{V}}\left\langle c_{-\vec{k}_{V}}c_{\vec{k}_{V}%
}\right\rangle -\xi_{0}\left\langle c_{0}\right\rangle +iP_{r},\label{eq2}\\
&  \frac{d\left\langle c_{-\vec{k}_{V}}c_{\vec{k}_{V}}\right\rangle }%
{dt}=-i(2\Delta\omega_{\pm\vec{k}_{V}})\left\langle c_{-\vec{k}_{V}}c_{\vec
{k}_{V}}\right\rangle -\nonumber\\
&  i\mathcal{D}_{\vec{k}_{V}}\left[  \left\langle c_{0}\right\rangle
(2\left\langle n_{\vec{k}_{V}}\right\rangle +1)\right]  -2\xi_{\vec{k}_{V}%
}\left\langle c_{-\vec{k}_{V}}c_{\vec{k}_{V}}\right\rangle ,\label{eq3}\\
&  \frac{d\left\langle n_{\vec{k}_{V}}\right\rangle }{dt}=2\mathcal{D}%
_{\vec{k}_{V}}\mathrm{Im}\left[  \langle c_{0}\rangle\langle c_{-\vec{k}_{V}%
}c_{\vec{k}_{V}}\rangle\right]  -\nonumber\\
&  2\xi_{\vec{k}_{V}}\left(  \langle n_{\vec{k}_{V}}\rangle-\bar{n}_{\vec
{k}_{V}}\right)  . \label{eq4}%
\end{align}
where $n_{\pm\vec{k}_{V}}=c_{\pm\vec{k}_{V}}^{\dag}c_{\pm\vec{k}_{V}}$ \cite{Lvov1944}. Here
we address the weak excitation regime in which we may disregard $\mathcal{H}%
^{(4MS)}$.

We first consider a valley-magnon pair with a $\delta$-function density of
states at $\omega_{0}/2$, which is justified for relatively thin disks as in
Ref. \cite{Makiuchi2023}. We analyze the decay dynamics of a non-equilibrium
state by switching off the external drive field $\mathcal{H}^{(d)}=0$. For
$t\ll1/(2\xi_{\vec{k}_{V}})$, $\langle n_{\vec{k}_{V}}\rangle(t)\sim\langle
n_{\vec{k}_{V}}(0)\rangle$ and Eq. (\ref{eq4}) does not contribute. The
solution is then $[\langle c_{0}\rangle,\langle c_{\vec{k}_{V}}c_{-\vec{k}%
_{V}}\rangle]^{T}=A_{+}e^{\lambda_{+}t}v_{+}+A_{-}e^{\lambda_{-}t}v_{-}$,
where $\lambda_{\pm}=\frac{1}{2}[\pm\sqrt{\mathcal{A}}-\xi_{0}-2\xi_{\vec
{k}_{V}}]$, $v_{\pm}=[-\xi_{0}-2\xi_{\vec{k}_{V}}\mp\sqrt{\mathcal{A}%
},-2i\mathcal{D}_{\vec{k}_{V}}(2\langle n_{\vec{k}_{V}}\rangle+1)]^{T}$, and
$\mathcal{A}=(\xi_{0}-2\xi_{\vec{k}_{V}})^{2}-4\mathcal{D}_{\vec{k}_{V}}%
^{2}(2\langle n_{\vec{k}_{V}}\rangle+1)$. For $\left\vert 4\mathcal{D}%
_{\vec{k}}^{2}(\langle n_{\vec{k}}\rangle+1)\right\vert \ll\left\vert (\xi
_{0}-2\xi_{\vec{k}})^{2}\right\vert $, $\lambda_{+}\approx-\xi_{0}$ and
$\lambda_{2}\approx-2\xi_{\vec{k}}$, \textit{i.e.} a \ fast ($\xi_{0}$) and
slow ($2\xi_{\pm\vec{k}}$) double exponential decay. When $4\mathcal{D}%
_{\pm\vec{k}}^{2}(2\langle n_{\pm\vec{k}}\rangle+1)>(\xi_{0}-2\xi_{\pm\vec{k}%
})^{2}$, the Kittel mode amplitude initially Rabi (cosine and sine) oscillates
with frequency $\sqrt{-\mathcal{A}}$ that is damped by $(\xi_{0}+2\xi_{\pm
\vec{k}_{V}})/2$, followed by a non-oscillatory decay $\sim2\xi_{\vec{k}_{v}}$.

\begin{figure}[ptb]
\includegraphics[width=0.5\textwidth]{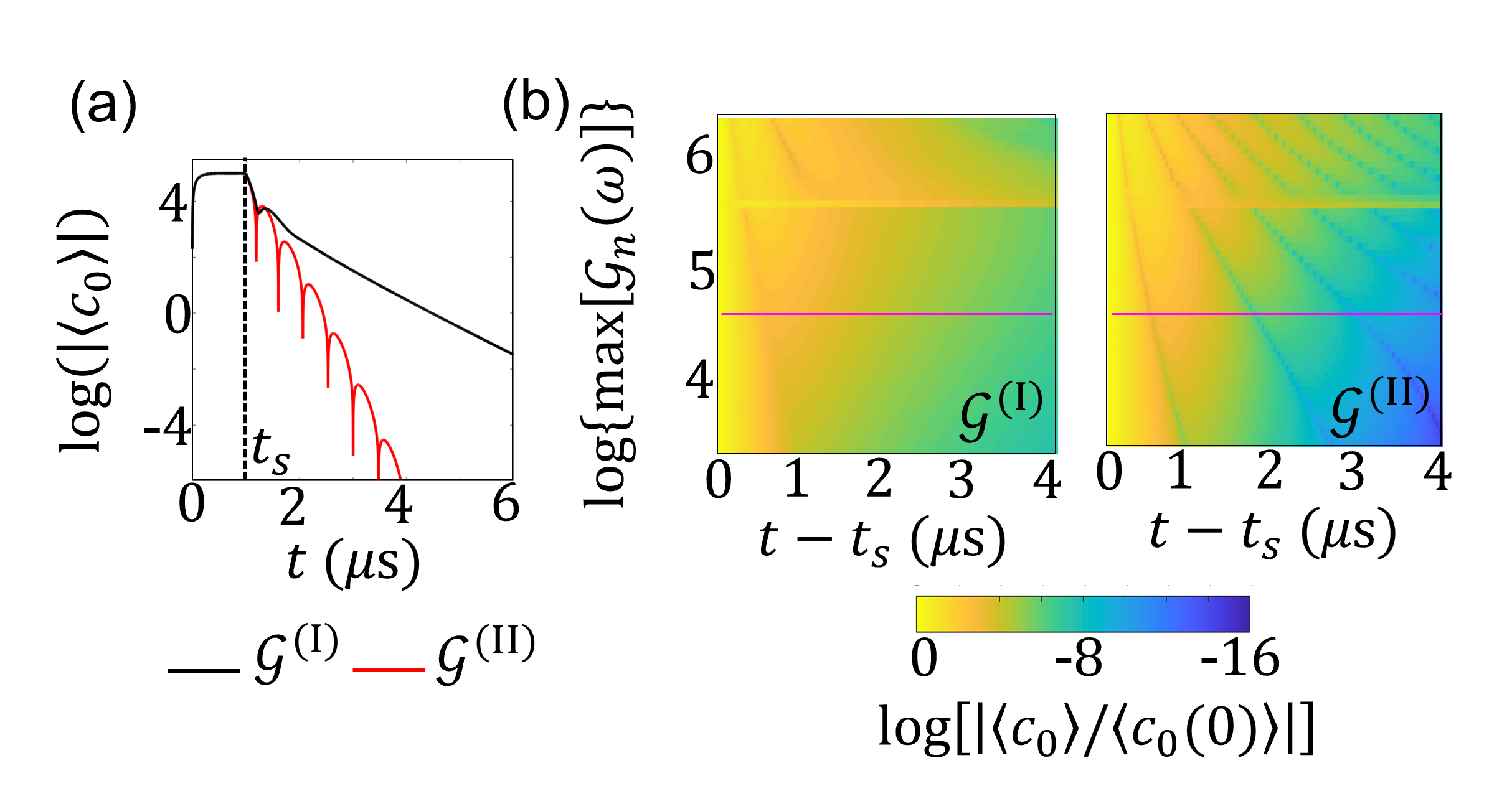}\caption{Decay dynamics for
a spectrum of valley magnons. (a) The time evolution for
$\mathcal{G}^{(\text{I})}=0.1$ and $\mathcal{G}^{(\text{II})}=0.1\Theta
(\omega-\omega_{0}/2)$. $P_{r}/\xi_{0}=10^{5}$. (b) The dependence of $\langle
c_{0}\rangle(t)$ calculated from Eq. (\ref{eq6}), on the peak amplitude of
$\mathcal{G}_{n} (\omega)$, for $\mathcal{G}^{(I)}$ (left panel) and $\mathcal{G}^{(II)}$
(right panel). The magneta lines in (b) correspond to $\langle n_{\omega}\rangle(t_{s})$ in
(a).}%
\label{fig3}%
\end{figure}

Hybridization of two Bosons (\textit{e.g.} magnons, photons, and phonons) with field
operators $a$ and $b$ is often pursued based on the \textquotedblleft beam
splitter\textquotedblright\ interaction $\mathcal{H}^{(BS)}=D(ab^{\dag
}+a^{\dag}b)$, which becomes strong at resonances, \textit{e.g}. in cavity
magnonics \cite{Rameshti2022} or standing sound waves in magnetic mechanical
resonators \cite{An2022}. The EOM of the Kittel mode $c_{0}$ and a boson $b$
with dissipation rate $2\xi_{\vec{k}_{V}}$ is the same as Eqs. (\ref{eq2})
and (\ref{eq3}) when $\langle n_{\vec{k}_{V}}\rangle=0$ and $\mathcal{D}%
_{\vec{k}_{V}}=D$. The differences between the nonlinear 3MS and the
hybridized systems arise from the non-Hermitian self-consistent effective
coupling that can be tuned via $\langle n_{\vec{k}_{V}}\rangle$, while $D$ is
a fixed parameter
\cite{Demidov2007,Demidov2008,Serga2013,Bozhko2016,Rezende2009,Rueckriegel2015}.

We show in the following that numerical solutions of Eq. (\ref{eq1})
corroborate the simple picture sketched above. In Figure \ref{fig2}(a) we show
results for resonant excitation in the time interval $t\in\left[
0,t_{s}\right]  $ with $P_{r}\neq0$. The decay of
$\langle c_{0}\rangle(t)$ with and without 3MS interaction ($\mathcal{D}%
_{\vec{k}_{v}}=0$ and $\bar{n}_{0}=\bar{n}_{\vec{k}_{v}}=0)$ is drastically
different for $t>t_{s},$ changing from fast ($\xi_{0}$) to slow ($2\xi
_{\vec{k}_{v}}$) even for a relatively weak $\mathcal{D}_{\vec{k}_{v}}\ll
(\xi_{0}-2\xi_{\vec{k}_{v}})/2$, as predicted by the simple model. Figure
\ref{fig2}(b) illustrates the effect of a relatively large 3MS interaction,
while maintaining $2\mathcal{D}_{\vec{k}_{v}}<\xi_{0}-2\xi_{\vec{k}_{v}}$ for
different thermal occupations of the valley magnons $\bar{n}_{\vec{k}}$ that
can be tuned by parametric pumping, the 4MS,\ and magnon-conserving phonon
scattering. When $\bar{n}_{\vec{k}}=0$, $\langle n_{\vec{k}}\rangle(t_{s})$
remains negligibly small and the dynamics is a fast decay followed by a slow
one [black line in Fig. \ref{fig2}(b)]. When we increase the average occupation of valley
magnons to $\bar{n}_{\vec{k}_{v}}=3.5\Theta(t-t_{s})$, where $\Theta(t-t_{s})$
is the Heaviside function, the 3MS$\ $becomes strong. When $\mathcal{D}%
_{\vec{k}_{v}}^{2}\left(  2\langle n_{\vec{k}_{v}}\rangle(t>t_{s})+1\right)
>(\xi_{0}-2\xi_{\vec{k}_{v}})^{2}$, the Rabi oscillation [red line in Fig.
\ref{fig2}(b)] is damped by $(\xi_{0}+2\xi_{\vec{k}_{v}})/2$. Adopting $\bar{n}_{\vec{k}_{v}}=3.5e^{-2\xi_{V}(t-t_{s})}\Theta(t-t_{s})$ which decays by $2\xi_{\vec{k}_{v}}$, leads to the
green line in Fig. \ref{fig2}(b) that initially decays like the strong coupling case but then by $2\xi_{\vec{k}_{V}}$.

Depending on the shape of the magnet and the applied magnetic field, the
Kittel mode may interact with a quasi-continuum rather than a discrete states
of valley magnons, such as in a relatively thick and wide film with
$\omega_{0}/2\geq\omega_{min}$ as in Fig. \ref{fig1}(b). This scenario is similar
to that of monochromatic photons interacting with an inhomogeneously broadened
ensemble of NV center spins
\cite{Putz2014,Putz2016,Krimer2015,Krimer2014,Julsgaard2013}. We extend our
theory by considering two limiting cases of $\mathcal{G}(\omega)=\mathcal{R}
(\omega)\bar{\mathcal{D}}^{2}(\omega)$, where $\mathcal{R}(\omega)$ is the
density of states (DOS) of valley magnons, $\bar{\mathcal{D}}(\omega)$ is the average of
$\mathcal{D}_{\vec{k}_{V}}$ at frequency $\omega$: (I) $\mathcal{G}^{(\mathrm{I}%
)}(\omega)=C\Theta(\omega-\omega_{0}/2)$, and (2) $\mathcal{G}^{(\mathrm{II}%
)}=C$, where $C$ is a constant. (I) and (II) correspond to $\omega_{min}=\omega_{0}/2$
and $\omega_{min}\ll\omega_{0}/2$, respectively. Figure \ref{fig3}(a) shows
the calculated dynamics for $C=0.1$, in which the decay for (I) is both fast
($\xi_{0}$) and slow ($2\xi_{\vec{k}_{v}}$), while for (II) it is only fast.
Next, we explain the dependence of the persistent coherence on the spectrum of
valley magnons in more detail.

\begin{figure}[ptb]
\includegraphics[width=0.5\textwidth]{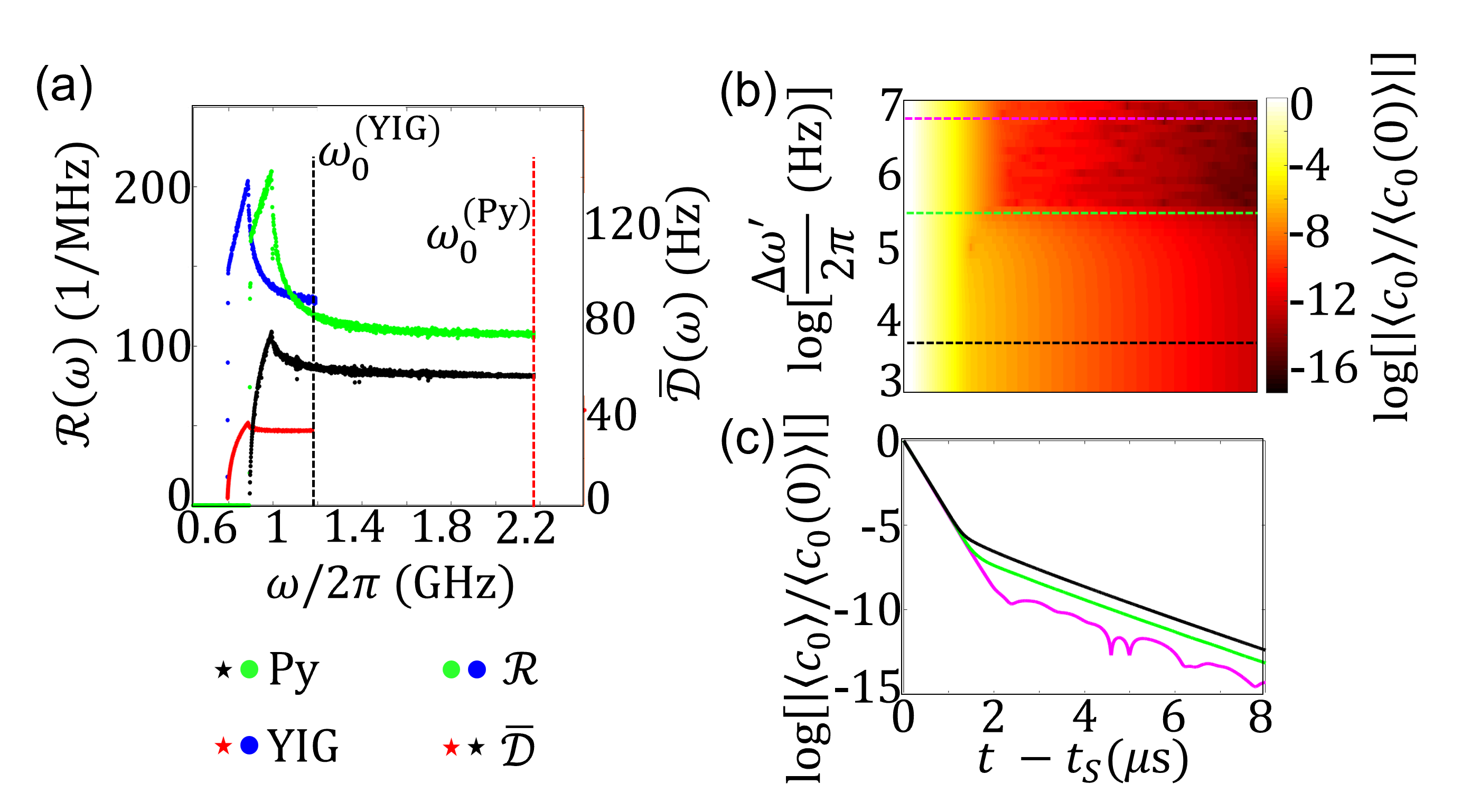}\caption{Calculated results
for a magnetic thin disk (thickness $d=1\,\mathrm{\mu}$m, radius
$100\,\mathrm{\mu}$m) at an external magnetic field $H=25\,$mT. (a) Density
of states $\mathcal{R}(\omega)$ and mean value of the 3MS interaction
coefficient $\bar{\mathcal{D}}$ up to the FMR\ frequencies $\omega_{0}^{YIG(Py)}$. (b)
The dependence of the decay $\log[\langle c_{0}\rangle(t)]$ on the detuning
$\Delta\omega^{\prime}=\omega_{0}/2-\omega_{min}$ for $P_{r}/\xi_{0}=10^{5}$.
(c) Time evolution for three $\Delta\omega^{\prime}$ (indicated by dashed
lines colored as in (b). In calculations, we used the spectrum of $0.1\,$GHz
width centered around $\omega_{0}/2$ and discretization of $0.01\,$MHz. The
oscillations in the relatively large $\Delta\omega^{\prime}$ are the artifact
of limited bandwidth and discretization.}%
\label{fig4}%
\end{figure}

The dynamics at $t>t_{s}$ is easy to understand when the valley magnons are initially
incoherent, \textit{i.e.} $\langle
c_{\vec{k}_{V}}c_{-\vec{k}_{V}}\rangle|_{\omega_{\vec{k}_{V}}=\omega}(t_{s})=0$. From
the full numerical calculations with $\langle c_{\vec{k}_{V}}c_{-\vec{k}_{V}%
}\rangle|_{\omega_{\vec{k}_{V}}=\omega}(t_{s})\neq0$ we conclude that coherent valley
magnons do not affect the time scales of interest\textit{.} Equations
(\ref{eq2})-(\ref{eq4}) then lead to
\begin{align}
&  \partial_{t}\langle c_{0}\rangle(t)=-\xi_{0}\langle c_{0}\rangle
(t)-\int_{0}^{\infty}\mathcal{G}(\omega)\left[  2\langle n_{\omega}%
\rangle+1\right]  d\omega\times\nonumber\\
&  \int_{0}^{t}d\tau e^{-i(2\Delta\omega-2i\xi_{\vec{k}_{V}})(t-\tau)}\langle
c_{0}\rangle(\tau), \label{eq5}%
\end{align}
for constant $\langle n_{\vec{k}_{V}}\rangle|_{\omega_{\vec{k}_{V}}=\omega
}=\langle n_{\omega}\rangle(t)=\langle n_{\omega}\rangle(0)$. The solution of
Eq. (\ref{eq5}) \cite{Krimer2014},
\begin{equation}
\langle c_{0}\rangle(t)=\langle c_{0}\rangle(0)\left[  \sum_{i}R_{i}+Q\right]
,\label{eq6}
\end{equation}
where%
\begin{equation}
Q=e^{-2\xi_{\vec{k}_{V}}t}\int_{0}^{\infty}\mathcal{G}_{n}(x/2)\mathcal{B}%
(x)e^{-ixt}dx,\label{eq7}
\end{equation}
$\mathcal{G}_{n}(x)=\mathcal{G}(x)[2\langle n_{x}\rangle+1]$, $R_{i}$ are the
residues of $e^{st}\mathcal{L}$ at the poles $s_{i}$ of $\mathcal{L}%
=\{s+\xi_{0}+\int_{0}^{\infty}d\omega\mathcal{G}_{n}(\omega)[s+i(2\Delta
\omega-2i\xi_{\vec{k}_{V}})]^{-1}\}^{-1}$, $\mathcal{B}(x)=\{[\pi
\mathcal{G}_{n}(x/2)/2]^{2}+[i(\xi_{0}-\xi_{\vec{k}_{V}})+x-\omega
_{0}+\mathcal{P}(x)/2]^{2}\}^{-1}$, and $\mathcal{P}(x)$ is the Cauchy
principle value of $\int_{0}^{\infty}\mathcal{G}_{n}(\omega^{\prime}%
/2)d\omega^{\prime}/(\omega^{\prime}-x)$.

$\langle n_{\omega}\rangle$ at $t=t_{s}$ in Fig. \ref{fig3}(a) is best fit by
a Gaussian of broadening $2\xi_{\vec{k}}$. We determine $\langle c_{0}\rangle(t)$
from Eq. (\ref{eq6}) by evaluating the poles $s_{i}$ and calculating the
integral in $Q$, for $\mathcal{G}^{(I)}$ and $\mathcal{G}^{(II)}$. In Fig.
\ref{fig3}(b), we assess the dependence of $\langle c_{0}\rangle(t)$ on the
peak amplitude of $\mathcal{G}_{n}$ which can be experimentally accessed by
varying $\langle n_{\omega}(t_{s})\rangle$. For $\mathcal{G}^{(I)}$, the
two-exponential dynamics persists even in the strong coupling regime, while
for $\mathcal{G}^{(II)}$, the decay rate remains $\gg2\xi_{\vec{k}_{V}}$
throughout when increasing $\text{max}[\mathcal{G}_{n}(\omega)]$ from
relatively small to large values. We thus confirm that an asymmetry of
$\mathcal{G}(\omega)$ with respect to and vicinity of $\omega_{0}/2$ is a necessary condition
for reaching limiting lifetime of $2\xi_{\vec{k}_{V}}$, \emph{i.e.} we should
stay close to the band edge.

The difference between the dynamics associated with $\mathcal{G}^{(I)}$ and
$\mathcal{G}^{(II)}$ can be illustrated by the spectrum $T(\omega_{t})=\langle
c_{0}\rangle(\omega_{t})/P_{r,t}$ which is the response to an excitation
$P_{r}=P_{r,t}e^{-i\omega_{t}t}$, while $\langle c_{0}\rangle(\omega_{t})$ is
the steady state $\langle c_{0}\rangle$ at frequency $\omega_{t}$, and
$\omega_{t}$ is in the frame of $\omega_{0}$. With $\left\vert P_{r,t}\right\vert
=1/\xi_{0}$, $T(\omega_{t})$ follows from Eqs. (\ref{eq2}) and (\ref{eq3}) for
constant $\left\langle n_{\vec{k}_{V}}\right\rangle $. When $\mathcal{D}%
_{\vec{k}_{V}}=0$, $\left\vert T(0)\right\vert =1$. When
$\mathcal{D}_{\vec{k}_{V}}\neq0$ and $\lim_{2\xi_{\vec{k}_{V}\rightarrow0}%
}\left\vert T(\omega_{t})\right\vert =1$ for some $\omega_{t}$, the Kittel
mode decays at the limiting rate of $2\xi_{\vec{k}}$. Straightforward algebra leads to $\lim_{2\xi_{\vec{k}_{V}\rightarrow0}}T(\omega
_{t})=i\xi_{0}\{[\omega_{t}+\mathcal{P}(\omega_{0}+\omega_{t})/2]+i[\xi
_{0}+\pi\mathcal{G}_{n}[(\omega_{0}+\omega_{t})/2]/2]\}^{-1}$. In case of
$\mathcal{G}^{(I)}$, $\mathrm{Im}[-iT(\omega_{t})^{-1}]>1$
$\forall\omega_{t}$, thereby $\vert T(\omega_{t})<1\vert$. On the other hand, when
$\omega_{t}<0$ for $\mathcal{G}^{(II)}$, we have $\mathcal{G}_{n}[(\omega
_{t}+\omega_{0})/2]=0$. Since $\lim_{\omega_{t}\rightarrow0^{-}}%
\mathcal{P}(\omega_{t}+\omega_{0})/2=\infty$ and $\lim_{\omega_{t}%
\rightarrow-\infty}\mathcal{P}(\omega_{t}+\omega_{0})/2=0$, $\omega
_{t}+\mathcal{P}(\omega_{t}+\omega_{0})/2=0$ for an $\omega_{t}<0$ at which
$T(\omega_{t})=1$ is reached. For partially suppressed density of states in $\mathcal{G}%
^{(II)}$ we therefore predict effects that resemble those of spectral hole
burning of inhomogeneously broadened NV center spin ensembles, an established
method to reach the dissipation rate of individual spins \cite{Krimer2015,Putz2016}.

Next, we address material dependence for samples of equal size. The 3MS coefficient $\mathcal{D}%
_{\pm\vec{k}}=\omega_{M}\mathfrak{g}_{k}\sin\theta_{\vec{k}}\cos\theta_{\vec{k}}%
(u_{\vec{k}}+v_{\vec{k}})(u_{\vec{k}}v_{0}+v_{\vec{k}}u_{0})/\sqrt{2S}$
\cite{Krivosik2010}, where $u_{\vec{k}}=\sqrt{(A_{\vec{k}}+\omega_{\vec{k}%
})/2\omega_{\vec{k}}}$ and $v_{\vec{k}}=-\text{sign}(B_{\vec{k}}%
)\sqrt{[(A_{\vec{k}}-\omega_{\vec{k}})/2\omega_{\vec{k}}]}$, $A_{\vec{k}%
}=\omega_{H}+\omega_{M}/2[2\alpha_{ex}k^{2}+\mathfrak{g}_{k}\sin^{2}\theta_{\vec{k}%
}+1-\mathfrak{g}_{k}]$, $B_{\vec{k}}=\omega_{M}/2[\mathfrak{g}_{\vec{k}}\sin^{2}\theta_{\vec{k}%
}+\mathfrak{g}_{k}-1]$, $\mathfrak{g}_{k}=1-[1-e^{-kd}]/kd$, $\alpha_{ex}=2A_{ex}/\mu_{0}M_{s}^{2}$,
$\theta_{\vec{k}}$ is the in-plane angle of $\vec{k}$ with magnetization,
$A_{ex}$ is the exchange stiffness, $M_{s}$ is the saturation magnetization,
$\omega_{M}=\mu_{0}\gamma M_{s}$, $\omega_{H}=\mu_{0}\gamma H$, and $\gamma$
is the gyromagnetic ratio, $S=VM_{s}/(\hbar\gamma)$ is the number of spins,
and $V$ is the sample volume. Since for valley magnons $u_{\vec{k}}\sim1$ and
$v_{\vec{k}}\sim0$, $\mathcal{D}_{\pm\vec{k}_{v}}\propto\sqrt{\omega_{M}}$.
From $\omega_{\vec{k}}=\sqrt{A_{\vec{k}}^{2}-|B_{\vec{k}}|^{2}}$ and some
simple algebra, we find the $k_{min}$ associated with $\omega_{min}$. The
approximation $k_{min}\approx1/\sqrt{4d\alpha_{ex}}$ is justified for
$\omega_{H}\gg\omega_{M}\alpha_{ex}k_{min}^{2}$. $\omega_{min}^{2}%
\approx\omega_{H}^{2}+5\omega_{M}^{2}\alpha_{ex}k_{min}/(4d)+\omega_{M}%
\omega_{H}[2\alpha_{ex}k_{min}^{2}+1/(k_{min}d)]$, while $\omega_{0}%
^{2}=\omega_{H}^{2}+\omega_{M}\omega_{H}$. $A_{ex}$ increases while
$\alpha_{ex}$ decreases with $M_{s}$. Therefore, $\omega_{min}$ increases with
$M_{s}$, but less so than $\omega_{0}$. Since the curvature $\partial
^{2}\omega_{k}/\partial{k^{2}}|_{k=k_{min}}$, and $\omega_{\vec{k}}%
^{2}-\omega_{min}^{2}\approx\omega_{H}\omega_{M}\theta_{\vec{k}}^{2}$ when
$\left\vert {\vec{k}}\right\vert {=k_{min}}$, increase with $M_{s}$, the DOS
of the valley magnons does not change much.

In Fig. \ref{fig4}(a) we plot the frequency dependence of the DOS
$\mathcal{R}(\omega)$, and average interaction strength $\bar{\mathcal{D}%
}(\omega)$ for a magnetic disk with thickness $d=1\,\mathrm{\mu}$m and radius
$r=100\,\mathrm{\mu}$m for typical parameters of YIG and permalloy (Py).
$\bar{\mathcal{D}}(\omega)$ vanishes at the band edge but then increases rapidly with
$\omega$. $M_{s}$ and $A_{ex}$ of Py are almost 10 times larger than those of
YIG. According to our analysis above, the DOS amplitude is about the same, but
$\bar{\mathcal{D}}$ increases substantially. We solve Eqs. (\ref{eq2})-(\ref{eq4}) for
the $\mathcal{R}(\omega)$ and $\bar{\mathcal{D}}(\omega)$ of YIG for different
values of $\Delta\omega^{\prime}=\omega_{0}/2-\omega_{min}$ which can be tuned
by the external field $H$ [see Fig. \ref{fig4}(a)]. For better comparison, the
drive is switched off at a time $t_{s}$ at which $\max[{\langle n_{\omega
}\rangle}]/S=10^{-10},$ chosen to be small enough that 4MS interactions may be
disregarded. Figures \ref{fig4}(b) and (c) show that the slow decay sets in
faster and the coherence persists much longer when $\Delta\omega^{\prime
}\rightarrow 0$. The relatively small (large) detuning corresponds to the case $\mathcal{G}^{(I)}$ ($\mathcal{G}^{(II)}$). Since $\bar{\mathcal{D}}$ is
larger for materials with larger $M_{s}$ such as Py, the persistent coherence
of YIG can be achieved with weaker drive, but it requires a larger $H$. For
sufficiently large lateral dimension $r>10\,\mathrm{\mu}$m, $\mathcal{R}%
(\omega)$ of the valley magnons does not change with $r$ while $\bar
{\mathcal{D}}(\omega)\propto1/r$. So, smaller disks require weaker excitation
for the same effect. Concluding, as long as $\xi_{0}\gg\xi_{\vec{k}},$ which
still has to be proven for Py, a substantial persistent coherence is easier to
achieve for Py than YIG, in spite of the larger Gilbert damping of the former.

\begin{figure}[ptb]
\includegraphics[width=0.5\textwidth]{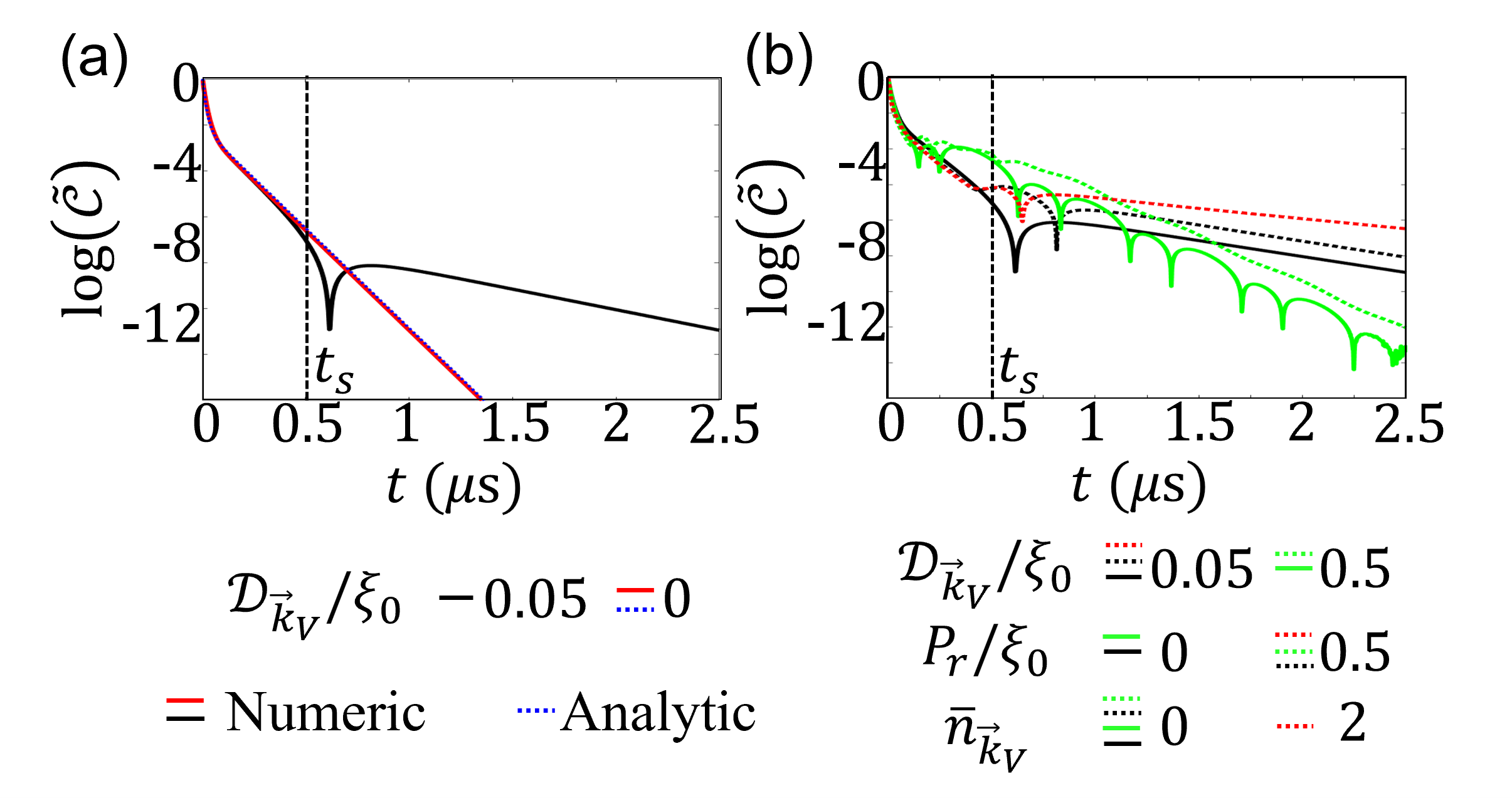}\caption{Decay of quantum
coherence. (a) $\mathcal{\tilde{C}}(t)$ as a function of time with a
relatively weak and completely without 3MS interactions. The blue dashed line
is our analytical result in the presence of 3MS. (b) $\mathcal{\tilde{C}}(t)$
with and without the resonant drive of the Kittel mode, now for weak and strong
3MS interactions. The red dashed line holds for $\bar{n}_{\vec{k}_{V}}=2$ for
weak 3MS and $P_{r}\neq0$. }%
\label{fig5}%
\end{figure}

\textit{Quantum coherence-} Here we address the decay of the quantum
interferences in a non-classical state such as a superposition of two coherent
states that may differ from that of coherent states and their statistical
superpositions. Here, we focus on the quantum coherence of an initial
non-classical superposition of Kittel mode coherent state. We adopt the
Schr\"{o}dinger cat state of the isolated Kittel mode as initial state,
$|\psi_{0}(0)\rangle=|\psi_{0,cat}\rangle=\mathcal{N}(|\alpha_{0}%
\rangle+|\alpha_{0}^{\prime}\rangle)$, where $|\alpha_{0}\rangle$ and
$|\alpha_{0}^{\prime}\rangle$ are two distinct coherent states with amplitudes
$\alpha_{0}$ and $\alpha_{0}^{\prime}$, respectively, and $\mathcal{N}$ is a
normalization constant. For simplicity and without loss of generality, we set
$\alpha_{0}^{\prime}=-\alpha_{0}$ and $\mathrm{Im}[\alpha_{0}]=0$. For a
Kittel mode in the absence of 3MS interaction, the density matrix in the
presence of a linear dissipation expressed by the $L_{\vec{k}}^{(L)}$ Lindblad
operators in Eq. (\ref{eq1}) can be exactly solved as $\rho(t)=\mathcal{N}%
^{2}(|\alpha_{0}(t)\rangle\langle\alpha_{0}(t)|+|-\alpha_{0}(t)\rangle\langle
-\alpha_{0}(t)|+[\mathcal{C}(t)|-\alpha_{0}\rangle\langle\alpha_{0}%
|+\mathrm{H.c}.])$, where $\alpha(t)=\alpha_{0}e^{-\xi_{0}t}$. $\mathcal{C}%
(t)=e^{-2\xi_{0}t}\exp[-4\alpha_{0}^{2}(1-e^{-\xi_{0}t})]$
\cite{Walls2008,Wiseman2010} determines the difference between quantum and
statistical superposition and$\ $vanishes faster than $|\alpha(t)|^{2}.$ At long times, the state becomes a completely statistical
rather than quantum superposition. Adding a resonant drive $P_{r}(c_{0}%
+c_{0}^{\dag})$, $\alpha(t)=\alpha_{0}e^{-\xi_{0}t}+(iP_{r}/\xi_{0}%
)(e^{-\xi_{0}t}-1)\ $does not affect $\mathcal{C}(t)$.

We now assess the effect of 3MS on the density matrix in the space of the
Kittel mode and a pair of valley magnons for two distinct initial conditions
$\rho_{cat}(0)=|\psi_{0,cat}\rangle\langle\psi_{0,cat}|\otimes\rho_{\pm\vec
{k}_{V},vac}$ and $\rho_{stat}(0)=|\psi_{0,stat}\rangle\langle\psi
_{0,stat}|\otimes\rho_{\pm\vec{k}_{V},vac}$, where `$vac$' indicates valley
magnon vacuum, $\left\vert \psi_{0,stat}\right\rangle \left\langle
\psi_{0,stat}\right\vert =\mathcal{N}^{\prime}(|\alpha_{0}\rangle\langle
\alpha_{0}|+|-\alpha_{0}\rangle\langle-\alpha_{0}|)$ is a statistical mixture
of two coherent states, and $\mathcal{N}^{\prime}$ is a normalizing constant. We label the density matrices evolving from these two
initial conditions $\rho_{cat}(t)$ and $\rho_{stat}(t)$ and integrate the
valley magnons out to obtain an effective density matrix of the Kittel mode
$\rho_{0,cat(stat)}$. For a measurement operator $\hat{O}$ with eigenvalues
$o$ and eigenstates $|o\rangle$, the coherence of an arbitrary state
$|\psi\rangle=\sum_{i}\mathfrak{c}_{i}|\phi_{i}\rangle$, where $\mathfrak{c}_{i}=\langle \psi|\phi_{i}\rangle$, emerges in the probability distribution $P(o)$ as $\sum_{i\neq j}\langle o|\phi_{i}\rangle\langle\phi
_{j}|o\rangle$, which vanishes when $\hat{O}$ is diagonal in $|\phi_{i}\rangle$.
In other words, the quantum interferences become observables for a measurement
operator that does not commute with the projection operator of the measured
states \cite{Walls2008}. Here the number operator serves this purpose via
$p_{n,cat(stat)}=\langle n|\rho_{0,cat(stat)}(t)|n\rangle$, where $|n\rangle$
is the $n$ number (Fock) state of the Kittel mode such that the time
dependence of $\mathcal{\tilde{C}}(t)=\sum_{n}[{p_{n,cat}-p_{n,stat}}]\left[
\neq\mathcal{C}(t)\right]$ measures the
decay of the quantum coherence \cite{Walls2008}. In the calculations below, we
use the same dissipation parameters as above.

Figure \ref{fig5}(a) shows our results for $\mathcal{\tilde{C}}\left(
t\right)  $ with and without (weak) 3MS and without drive $P_{r}=0$, compared
with the $\mathcal{C}(t)$ (blue dashed line) for equal starting amplitudes
$\mathcal{C}(0)=\mathcal{\tilde{C}}(0)$\textit{.} We find that 3MS increases
also the quantum coherence, similar to the effect of the beam-splitter
interaction of hybrid systems. As discussed above, 3MS introduces a $\langle
n_{\vec{k}_{V}}\rangle$ dependence, however, that may be tuned indirectly by driving the
Kittel mode or directly occupying the valley magnons $\bar{n}_{\vec{k}_{V}}$. Figure
\ref{fig5}(b) shows $\mathcal{\tilde{C}}(t)$ when $P_{r}=0$ and $P_{r}%
=0.5\xi_{0}$, for a weak $\mathcal{D}_{\vec{k}_{V}}/\xi_{0}=0.05$ and a strong
$\mathcal{D}_{\vec{k}_{V}}/\xi_{0}=0.5$ 3MS. The enhanced quantum coherence is
evident. In Fig. \ref{fig5}(b), we demonstrate that the persistent coherence
may be further enhanced when both $P_{r}=0.5\xi_{0}$ and $\bar{n}_{\vec{k}%
_{V}}=2$. The decay of the quantum coherence  $\mathcal{\tilde{C}}(t)$ limits
the visibility of the negativity of the Wigner function of the Kittel mode in
the tomography \cite{Hioki2021,Makiuchi2023} at low temperatures, but
according to our calculations the difference with the classical decay rates is minor.

\textit{Conclusion-} We refine the model calculations of the interaction of
the Kittel mode with valley magnons that enhances the magnetic coherence in
extended films and microstructures. In spite of the rather weak nonlinear
interactions in the samples and microwave powers explored to date, a
substantial persistent coherence has been reported \cite{Makiuchi2023}. We
uncover the role of the spectral dependence of three magnon interactions,
density of states, and applied magnetic fields and predict that the decay
rates of quantum and classical coherence are of the same order of magnitude.
Our model holds also for materials such as Py, and for other interacting three-level systems with mode-dependent dissipation rates
\cite{Korber2020}, such as heterostructures composed of different magnetic
materials \cite{Sheng2023} or nitrogen-vacancy centers in diamond
\cite{Carmiggelt2023}. Other interesting platforms are antiferromagnets such
as $\text{MnF}_{2}$ with hyperfine interactions that generate Suhl-Nakamura-de
Gennes nuclear spin waves \cite{Suhl1958,Nakamura1958,DeGennes1963,Shiomi2019}%
. The three-field resonance of electron spin waves, nuclei spin waves, and
phonons was experimentally demonstrated long ago
\cite{Hinderks1968,Andrienko1991} and can now be modeled in a modern fashion
and implemented for enhancing the lifetime of electron magnons.

\textit{Acknowledgments-} We acknowledge support by JSPS KAKENHI (Grants No.
19H05600, 19H00645, 21K13847, 21K13886, 22H04965, 22K14584), JST CREST (Grant No. JPMJCR20C1),  JST PRESTO (Grant No. JPMJPR20LB), Advanced Technology Institute
Research Grants 2022, and partially supported by the Institute for AI and Beyond of the University of Tokyo, and the
IBM-UTokyo lab.


\begin{thebibliography}{99}
\bibitem {SM}See supporting material [URL will be inserted by publisher] for
details of the analysis of a Kittel mode interacting with a continuum
of valley magnons.
\end{thebibliography}
\end{document}